\documentstyle[epsfig,proceedings]{crckapb}

\newcommand{\beq}{\begin{equation}}
\newcommand{\eeq}{\end{equation}}
\newcommand{\bd}{\begin{displaymath}}
\newcommand{\ed}{\end{displaymath}}

\newcommand{\gapprox}{\;\rlap{\lower 2.5pt
             \hbox{$\sim$}}\raise 1.5pt\hbox{$>$}\;}
\newcommand{\lapprox}{\;\rlap{\lower 2.5pt
             \hbox{$\sim$}}\raise 1.5pt\hbox{$<$}\;}
\newcommand{\bfg}[1]{\setbox0=\hbox{#1}%
  \kern-.025em\copy0\kern-\wd0
  \kern.05em\copy0\kern-\wd0
  \kern-.025em\raise.0433em\box0}

\begin{opening}
  \title{Theory of solar luminosity variations}  
  \author{H.C. Spruit}
  \institute{Max-Planck-Institut f\"ur Astrophysik\\ Postfach 1523, D-85740 
Garching, Germany}
\end{opening}
\begin{document}
\begin{abstract}
The theory of stellar structure can be used to identify the most plausible
mechanisms for the irradiance variations associated with the solar cycle. Changes in surface emissivity, i.e. the reduced cooling in spots and enhanced emission by small scale magnetic fields, are the most effective mechanisms and account for most of the observed variation. Helioseismology will soon be able to test the consequences of changes in surface emissivity, and distinguish them from other scenarios for irradiance variability.

\end{abstract}

\section{Introduction}

When a sunspot group appears on the surface of the Sun, it reduce the brightness 
locally. Does this lead to an actual reduction of the Sun's brightness as 
measured on earth (the so-called irradiance)? Or is the `missing flux' 
re-emitted 
elsewhere on the solar surface? Faculae are bright areas, visible especially 
near the limb of the Sun. Do they contribute to the irradiance? Does the  
irradiance vary only because of spots and faculae, or are there other effects as 
well, and can these be large enough to influence climate on earth? 

Some of these questions have been answered by actual measurements with 
space-base radiometers over the past 20 years (figure 1). The data show a clear 
variation, obviously correlated with the solar cycle: the solar magnetic field 
does indeed cause the Sun's brightness to vary a bit, by about 0.1\% between 
minimum and maximum of the cycle. The record shows prominent sharp dips; 
these are associated with the passage of individual sunspot groups across the 
disk. When a large group appears, the Sun actually does become dimmer by a tiny 
bit. The smoothed data (thick line in the figure) shows, however, that the 
{\it average} trend is the opposite: at maximum activity, the Sun is brighter. 

The level of variation seen in figure 1 is not large, and by general opinion in the 
climate modeling community, insufficient to cause variations in the earth's 
climate, by a factor of 10 or so (see the contributions in Nesme-Ribes et al., 
1994, Friis-Christensen et al. 2000). Even if the Sun were to settle into a 
state of activity as seen during the minima in figure 1, the average reduction of 
irradiance by 0.05\% would have negligible effects, and probably would not 
explain events 
like the so-called `little ace age'\footnote{A period during the 17th century 
when the climate in Europe was colder by a degree C or so. This cold spell 
appears to have been limited mostly to Western Europe, however.}.

\begin{figure}
\centerline{\psfig{file=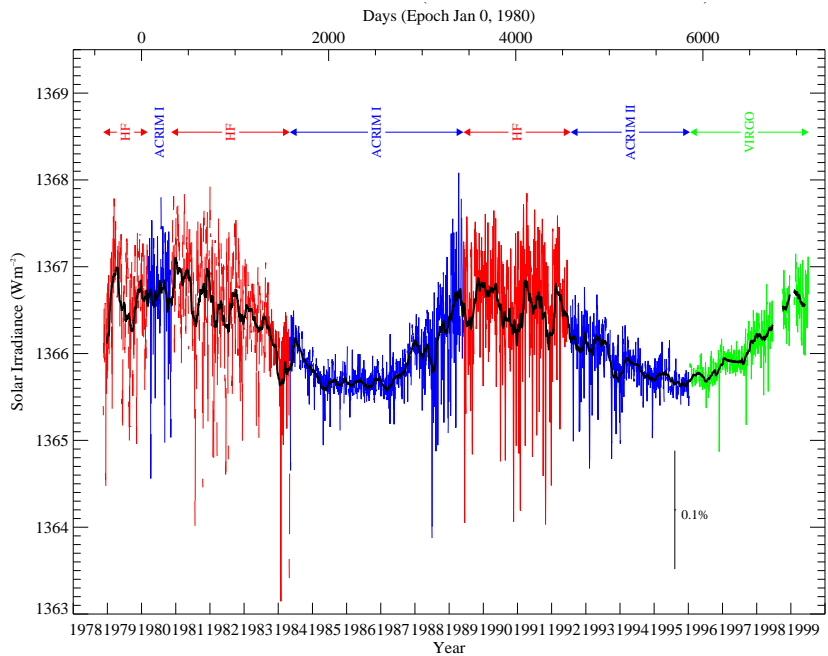,width=0.71\textwidth,angle=90,clip=}}
\caption{Solar irradiance record, showing the variation due to the solar cycle. 
From Fr\"ohlich and Lean, 1998 (see also {\tt 
http://www.pmodwrc.ch/solar\_irr/sola\_irr.html})}
\end{figure}

For climate modeling, the record of figure 1 is too short. One would like to know 
what the irradiance has been in the past, during the last 10000 years, for 
example, so comparisons can be made with climate records such as have been 
obtained from ice cores in Greenland and the Antarctic (see the contributions 
Friis-Christensen et al. 2000). Is it conceivable, 
for example, that figure 1 shows only a fraction of what the solar magnetic 
field is capable of doing, and that variations of, say, 10 time larger amplitude 
could have been caused by the solar magnetic field, or something else that has 
not yet been discovered on the Sun? These are questions that can excite the 
imagination, including that of funding agencies. 

In the absence of reliable independent indicators of the past brightness of the 
Sun, it is of some importance to understand how the variation seen in figure 1 
come about. If theoretical understanding is sufficient, and accepted by the 
community, it may be possible to assess with more certainty whether 
irradiance variations larger than those seen so far are likely to have occurred
in the past, and perhaps influence the climate.

Until the advent of helioseismology (see Dziembowski, elsewhere in this 
volume), our knowledge of the internal structure of the Sun was based on the 
theory of stellar structure and evolution. Since this theory is reasonably 
simple, confidence in the models was high even before helioseismological tests 
became possible. The main uncertainty had to do with the exact values of the 
opacity of stellar plasma. Increasingly accurate values for this were 
computed from first principles (atomic physics) by extensive numerical work. 
The basic correctness of the theory was then confirmed by helioseismology, to 
such an extent even that one is now confident that the `solar neutrino problem' 
has its cause in neutrino physics, not inaccuracies in the solar models (e.g. 
Bahcall 1996, Richard et al. 1996, Brun et al. 1998).

Though these solar models are spherical and only slowly evolving in time, the 
theory behind them also apply to time dependent and non-spherical 
disturbances. In this form, the theory makes definite predictions about the 
thermal behavior of the Sun, that is, the evolution in time and position in the 
star of perturbations that might be related to the observed irradiance 
variations. In the following I present my view how this theory can be used to 
locate the most plausible source of irradiance variations. After this, I discuss 
how the new heliosmological measurements currently being made can be used to 
test the predictions of proposed mechanisms of irradiance variation.

Conceptually, one can separate the thermal effects of magnetic fields
into three types:

1. `Sources and sinks'. The generation of a magnetic field involves the
conversion of energy of motion into magnetic energy. Since the motions
in the solar envelope are thermally driven, this ultimately means conversion
of thermal into magnetic energy: building up a magnetic field produces a
thermal sink somewhere. The opposite happens when the field decays: magnetic
energy is converted into heat. These thermal effects exist only during {\it
changes} in the magnetic energy content of the envelope.

2. `Shadows', or changes in the heat transport coefficient. Magnetic fields 
interfere with convection, causing a reduction in the efficiency of heat 
transport in the envelope. In contrast to (1), these changes last as long as the 
magnetic field itself is present.

Both these sources and sinks, and changes in the transport coefficient cause
thermal perturbations, varying with magnetic activity, which propagate through
the envelope and cause variations in surface energy flux.
Related to the second class of per\-tur\-ba\-tions are:

3. The effects of magnetic fields at the surface of the star. Sunspots, being 
dark, radiate less than the surrounding photosphere, while the small elements 
that make up plages and the network have an excess emission. In addition, it is 
conceivable that the magnetic elements have an indirect effect surface by 
modifying the convective flow in their surroundings slightly (for which there is 
some observational evidence, see section \ref{magmod}).

\section{Time scales}
The response of the Solar envelope to thermal perturbations is not governed by a 
single time scale, but by a wide range of time scales. The longest of these is 
the thermal time scale of the Sun as a whole, called the Kelvin-Helmholtz time
scale. It is about $10^7$ years, namely the thermal energy content of the Sun, of 
the order $10^{41}$ Joule, divided by its luminosity of $4\,10^{26}$ W. Hence, on 
time scales of human interest, the Sun has an  extremely large thermal inertia. 
If the central heat source of the Sun were switched off, the internal structure 
and the luminosity would start to change only on this long time scale. 

The Kelvin-Helmhotz time scale is a global time scale for the Sun as a whole; 
more generally, we can define the {\it thermal time scale} $\tau_{\rm t}$ as a 
function of depth:
\beq
\tau_{\rm t}(z) \equiv U(z)/L(z)\approx{1\over L}\int_{R-z}^R4\pi r^2 u {\rm 
d}r,
\label{tt} 
\eeq
where $L$ is the luminosity at depth $z$, $U$ the thermal energy of the
envelope down to a depth $z$, and $u$ the thermal energy per unit volume,
approximately (for an ideal gas of constant $\gamma$) given by  $u= 
{P/(\gamma-1)}$. This is the time scale on which the structure of the envelope, 
and the observed luminosity, would start changing when the heat flux in the star 
were interrupted, by some magical means, at depth $z$. Some rough values for 
this quantity are $\tau_{\rm t}\sim 10^5$yr at $z=2\,10^5$km (depth of the 
convection zone), 10 yr at $20\,000$km (the size of a supergranule), 10 hrs at 
$2000$km (size of a granule). This shows that the thermal time scale depends 
rather strongly on depth in the Sun. As a result, the thermal response of the Sun 
also depends critically on the location of the disturbance. Near-surface 
disturbances have much larger effects than ones orginating near the base of the 
convection zone.

A second kind of time scale involved in thermal readjustments is the {\it
diffusive} time scale. This is the time scale on which differences in entropy 
between different parts of the convection zone are ironed out by the convective 
flows. This is easily computed in a `mean field' or diffusion model for 
convection, in which the local convective heat flux is assumed to be a function 
of only the local conditions (thermodynamic state of the gas and the 
temperature gradient). The validity of this model is questionable, 
since it is now known that stellar envelope convection is in fact extremely 
non-local, with almost all flows generated by cooling at the stellar surface (e.g. 
Nordlund 1986, Stein and Nordlund 1989, for reviews see Nordlund and Stein 
1996, Spruit 1997). The diffusion model has the advantage, however, that it is  
simple to apply, and is (unfortunately) still almost universally accepted.  For 
these 
historical and practical reasons, I still use it for the estimates in this section. 
Estimates in the context of the new nonlocal view of envelope convection have 
not yet been developed to the same level of detail. I return to this in section 
\ref{new}, where the  examples of heat flux blocking by sunspots and `thermal 
shadows' are considered in the nonlocal convection picture. I argue there that 
the new nonlocal view only strengthens the conclusions based on the diffusion 
model. 

For small perturbations, the heat flux in the local model is proportional to 
the entropy gradient, and time dependent perturbations are governed by a 
diffusion equation. The `turbulent' diffusion coefficient in this equation is 
$\kappa_{\rm t}\approx {1\over 3} l_{\rm c} v_{\rm c}$ where $l_{\rm c}$ and 
$v_{\rm c}$ are the convective length scale and velocity. In mixing length 
models of the solar convective envelope, this quantity varies only weakly, at a 
value of the order $10^{13}$ cm$^2$/s. In this model thermal inhomogeneities 
(more precisely: entropy inhomogeneities) of length scale $d$ are smoothed by 
turbulent diffusion on a time scale
\beq \tau_{\rm d}=d^2/\kappa_{\rm t}. \label{dt}\eeq 
For $d=2\,10^5$km this is about 1 yr, for $d=2000$km about 1 hr. Comparing
$\tau_{\rm t}$ and $\tau_{\rm d}$, we see that they are of similar magnitude 
close to the surface (to be precise: in the surface boundary layer where 
convection is not efficient enough to keep the stratification close to adiabatic). 
In deeper layers, the thermal time scale is much {\it longer} than the diffusive 
time scale, by a factor of up to $10^5$. 

The two time scales measure different types of thermal adjustment process. 
These same processes appear in the thermal behavior of, say, a chunk of metal, 
aluminum for example, heated from the inside and suspended in space. The 
thermal time 
scale is the time scale on which its temperature adjusts to a change in the heat 
input, such that the heat radiated from the
surface into space balances the heat input again. It is determined by the
heat capacity [$U$ in eq. (\ref{tt})] and the power level ($L$). The time scale
on which different parts of the chunk equilibrate to the same temperature is
governed by a different process, namely thermal conduction (the equivalent
of the turbulent diffusion in the Sun). The diffusion (conduction) time scale
is much shorter than the thermal time scale, because of the large heat 
conductivity of Al. In the Sun, it is the very large turbulent diffusivity in the 
bulk of the convection zone that causes the very short diffusive time scale 
compared with the thermal time scale.

How do these different time scales come into play when the convection zone is
thermally perturbed by, say, the storage of energy in a growing magnetic
field? Such perturbations can be computed in detail, either by numerical
methods (Endal et al. 1985) or more analytically. We can, for example, consider
the initial value problem in which a perturbation is allowed to
evolve in time by heat transport in the convection zone. In general this
evolution has components on all the time scales of the problem, including the
very long thermal time scale. Detailed calculations of this problem are 
discussed
briefly in the next subsection (for a more complete analysis, see Spruit,
1982ab, 1991, Arendt, 1992). 

\subsection{Modes of thermal relaxation}
In the mixing length approximation for convective energy transport the heat flux 
is given by:
\beq
{\bf F}=-\kappa_{\rm t}\rho T\nabla S, \label{ml}
\eeq
where ${\bf F}$ is the heat flux, $\kappa_{\rm t}$ the turbulent diffusivity, 
$\rho$
the density, $T$ the temperature and $S$ the entropy. The energy equation, in
the absence of sources, is given in terms of the entropy by
\beq
\rho T {{\rm d}S\over {\rm d}t}=-{\rm div}{\bf F}.\label{en}
\eeq
As an example, consider the one dimensional problem in which all quantities
depend on depth $z$ only. In this case, combining (\ref{ml}) and
(\ref{en}), and neglecting flows we get
\beq 
{\partial S\over\partial t} = {F\over \rho T H} + \kappa {\partial S\over
\partial z^2}, \label{ts}
\eeq
where $H={\partial{\ln\rho T}/\partial z}^{-1}$ is the pressure scale height.
For perturbations in which the left hand side is balanced by the first term
on the right, the time scale is of the order $F/(c_p\rho T H)$, which is just
a thermal time scale (cf. \ref{tt}). For modes in which the left
hand side is balanced by the second term, we evidently have an ordinary
diffusion problem, hence these modes evolve on diffusive time scales. The 
first term has appeared as a consequence of the
stratification of pressure, ie. the inhomogeneity of the convection zone.
In a homogeneous medium (as in a chunk of Al), the thermal time scale comes
in only through the radiating surface boundary condition (which one could also
consider as a form of inhomogeneity). If $T_s$ is the surface temperature, this
condition is 
\beq \sigma T_s^4=F. \label{su}\eeq
Since the perturbations in entropy outside spots and the magnetic elements
making up the small scale field are small, the reponse of the convection zone
can be computed by linearizing eqs. (\ref{ml}),(\ref{en}),(\ref{su}). An
arbitrary initial perturbation may then be decomposed in the standard way in
terms of the eigenmodes of the problem. These are found by setting $\delta S =
e^{\eta t}f({\bf r})$, and solving for $f({\bf r})$ with the decay rate $\eta$
as eigenvalue (the problem thus defined has only decaying solutions).
Specializing to the one dimensional case ($f(z)$) one finds a series of modes
ordered by the number of nodes $n$ of $f$. The fundamental ($n=0$) has a decay
rate of the order $\eta_0\sim\tau_t^{-1} $ that is, this mode decays on the
thermal time scale. The higher modes decay  with rates of the order
$\eta_n\sim n^2\tau_d^{-1} $, they decay on diffusive time scales.

The extreme separation of time scales, $10^5$yr for the thermal time scale and 
1yr for the longest diffusive time scale (with $n=1$) leads to a simple picture. 
The high 
heat conductivity tends to keep the convection zone entropy uniform, so that 
sources and sinks of energy are shared by a large part of the convection zone.
This large part of the convection zone has a large heat capacity, however, so
that sources and sinks tend to spend their efforts in slow heatings and coolings 
of large amounts of mass, rather than propagating to the surface.

How do these different time scales come into play when the convection zone is
thermally perturbed by, say, the storage of energy in a growing magnetic
field? The evolution of such perturbations can be computed in detail, either by 
numerical
methods (Endal et al.~1985, Gilliland, 1988) or more analytically. We can, for
example, consider the initial value problem in which a perturbation is allowed
to evolve in time by thermal transport in the convection zone. In general this
evolution has components on all the time scales of the problem, including the
very long thermal time scale. Formal aspects of this problem have been 
discussed elsewhere (Spruit 1982ab, 1991, Arendt 1992). In the following,
the basic conclusions of these analyses are summarized.

\section{Expected level of luminosity variations}
For quantitative estimates, the strength of the field and its filling factor in the
convection zone have to be specified. Assume that we have a layer of field with
strength of the order of $10\,000$G (equipartition with the convective flows as
estimated by a mixing length model), one scale height deep, near the base of the
convection zone (where most of the magnetic flux is probably located). I 
summarize here some results, discussed in greater detail elsewhere (Spruit 
1991).

\subsection{sources-and-sinks}
If the energy needed to build up the solar magnetic field during one half of the 
solar cycle is taken out of the thermal energy near the base of the convection 
zone, the calculations show that a surface luminosity variation of only 
$\delta L/L\sim 10^{-7}$ results. This is due to the very large heat capacity of 
the lower 
convection zone. The effect is stronger if the source of the magnetic field is 
assumed to be closer to the surface, but is still much smaller than the observed 
effect. Recent models for the emergence of magnetic flux from the base of the 
convection zone (D'Silva and  Choudhuri 1993, D'Silva and Howard 1993, Caligari 
et al.~1998) indicate that the actual field strength at the base of the 
convection zone is probably about 10  times higher than the equipartition 
estimate, of the order $10^5$ G. The magnetic energy per unit of  magnetic flux 
is then also 10 times 
higher, but this still does not lead to a significant luminosity effect.

\subsection{shadows}
If magnetic fields interfere with convection, a thermal perturbation develops 
as well. If at some depth below the surface there is an area where magnetic 
fields reduce the convective flows, the surroundings will be affected thermally, 
and one might hope to see a `thermal shadow' at the surface.  This can again be 
computed in the diffusion assumption for convection (Spruit, 1977). The 
amplitude of the effect depends crucially on the `covering factor'. If a reduction
of convective efficiency is assumed that uniformly covers a horizontal surface
at some depth $z$, magnetic fields comparable to equipartition with convective 
flows can have stronger effects than the source-and-sink perturbations 
(Gilliland, 1988). A measurable effect, however, is predicted only if the field 
is located close below the surface. At face value, this would make changes in 
convective efficiency a reasonable candidate for irradiance variations. 

In practice, the effect is very strongly reduced, however, if the change in 
convective efficiency does not extend over an entire horizontal surface. If there 
are `holes' in this cover, the effect is much smaller. This is because the 
turbulent heat conductivity in the
convection zone is so high that the heat flux is easily `shunted' past blocking
objects below the surface (Spruit, 1977). Compare this with efforts to 
thermally insulate homes. To be effective, it is not enough that {\it most} of the 
building's outer surface is very well insulated. A more modest insulation but 
without any leaks is more effective. For this reason, thermal shadows of 
subsurface fields are probably not very important (see also section 
\ref{shad:new}).

\subsection{surface effects}
\label{sef}
By far the most effective way in which a magnetic field influences the 
irradiance is by its effect on the {\it net surface emissivity}. The reduced 
emission from a sunspot area shows up directly in the irradiance records as a 
dip tracking the passage of the area across the disk. The only complication is 
that one might expect, depending on one's prejudices, that part of this reduction 
could be compensated by a brightening elsewhere. For example in the form of a 
`bright ring' surrounding the spot. Evidence of such bright rings is absent for 
most spots that have been studied for this effect, and where observation of a 
ring is reported, it compensates for only a small fraction of the spot blocking
(e.g.\ Rast et al., 1999).  Brightening of the photosphere around a spot is hard to 
measure because it is usually masked by
facular emission, which is a different effect. Such faculae are part of all active 
regions. Though their excess emission is rougly of the same order as the spot 
deficit, this is a coincidence that does not hold any more on closer inspection. 
For example, they are present in the same amount in active regions with or 
without sunspots, and their contribution to the irradiance variation does not 
correlate very well with  spot areas (Foukal and Lean 1986, Lean et al. 1998).

The absence of bright rings is understood in terms of a turbulent diffusion 
model for the heat flux in the convection zone (Spruit 1977, 1982b, Foukal et 
al.~1983, Chiang et al. 1984). The `blocked heat flux', for the most part, does not 
reappear elsewhere on the surface, but stays inside the convection zone, being 
stored/released on the very long thermal time scale of the convective envelope.
This conclusion holds, in the diffusion model, as long as the blocking effect of
the spot extends to a depth of at least 1000 km, a mild requirement given that 
the observed Wilson depression of the umbra of a spot is already of the order of
500km.

\subsection{spot blocking: stove top analogy}
The absence of bright rings in the diffusion model can be understood by a 
kitchen analogy (for an improved, but more qualitative, model based on the new 
nonlocal view of convection, see section \ref{spot:new}). Imagine a stove top 
with an electric heater plate. Assume this plate to be made of a massive block 
of a metal with a high thermal conductivity, Aluminum or Copper, say. Such a 
block, like the solar convection zone, has two time scales, the thermal time 
scale, given by the heat content divided by the heating rate, and the thermal 
conduction time scale given by the thermal diffusion time across the block. 
These are different physical quantities and their values can be quite different. 
If the thermal conductivity is large, the diffusion time scale can be much 
shorter than the thermal time scale. In the solar convection zone this is the 
same except that the ratio of time scales ($\sim 10^5$) is much more extreme. 

Suppose the heater plate has been on and has reached a steady temperature, so 
that heat input balances the conduction/convection losses at the top. Suppose 
also that the ratio of time scales is sufficiently large. Then put a piece of 
thermally insulating material on top, covering part of the surface of the plate 
(as a model for the effect of a spot). Where does the blocked heat flux go? The 
answer is of course that it does not go anywhere, the blocked heat stays in the 
block, heating it up slowly, instead of being radiated at some other part of the 
surface. On account of the large thermal conductivity, the temperature in the 
block stays uniform, and almost no `bright ring' appears around the spot. When 
the insulating spot is removed after a while, the heat flux returns to its old 
value instantaneously. The temperature has increased only slightly during the 
blocking phase, and this slight increase disappears again on a thermal time 
scale.

\subsection{excess emission by small scale magnetic fields}
Apart from the highly visible sunspots, the Sun has a small scale magnetic field 
consisting of a very large number of tiny magnetic elements. They are called 
faculae if their density is high, and `network' at lower density, and collectively 
contain much more magnetic flux than sunspots. They are visible especially near 
the solar limb, showing that they radiate more isotropically than the 
limb-darkened normal solar atmosphere. 

This excess emission has two 
components, one originating in somewhat higher atmospheric layers, in the form 
of line- and UV-continuum emission. This component probably is due to 
dissipation of mechanical (wave) energy carried by the magnetic field. In 
addition, there is a component seen in white light, originating in the surface 
layers of the magnetic structures. This second component is understood 
theoretically as a radiative transfer effect (Spruit 1977): the radiating surface 
in a magnetic element is a little `dimple' in the surface. The large number of 
small magnetic elements effectively corrugates the surface, and makes it 
radiate both more effectively and more isotropically, just like a rough surface 
has a higher emissivity than a polished surface. This effect increases the 
radiative flux from any part of the surface where small scale magnetic fields 
are present. It is a net effect (predicted before observation, Spruit 1977) that 
is not balanced by a deficit somewhere else on the surface. Since the magnetic 
elements are so small, they are hardly resolved in the observations. Recent 
high-resolution observations (Topka et al. 1997), however, verify the picture 
predicted by flux tube models (Spruit 1976 figure 11,  Steiner et al. 1997) for 
the appearance of small magnetic elements.

Since the effect of small scale fields is an increased {\em emissivity}, it 
increases the {\em cooling rate} at the surface.

\subsection{changes in granulation in active regions}
\label{magmod}
The changes of surface emissivity by magnetic fields discussed above are due to 
the different radiative properties of small magnetic elements (`flux tubes'). 
More indirect effects may play a role as well. The shape of granules appears 
to be different in magnetic regions (Macris and Roesch 1983, Muller and Roudier 
1984, Muller 1986, Title et al. 1992). They are smaller, more irregular, and the 
measured flow speeds are lower. These effects are present {\it in addition} to 
the  magnetic elements themselves, which also change the appearance of 
granulation by filling in the intergranular lanes, especially when seen at low 
spatial resolution. The changes may be the result of the geometric constraints 
the magnetic tubes put on the convective flow outside them. 

Since the flow appears to be different, it would seem possible that the heat 
flux it carries is also different. This might contribute to the observed solar 
cycle variations of irradiance (Muller 1986, Kuhn et al. 1988). Limits on 
this contribution can be put by observations of the {\it colors} of the solar 
cycle variation signal. Solanki and Unruh (1998) find that 
the wavelength dependence of the irradiance signal argues against a dominant 
contribution from small changes of surface temperature over a larger areas, but 
do not quote quantitative limits.  In order to contribute in the right sense to the 
solar cycle variation, the constraints imposed on the flow by the magnetic flux 
elements would have to lead to an {\it increase} in the heat flux carried by 
granulation. Direct (spatially resolved) measurements of a heat flux change in 
granulation in magnetic regions are probably difficult, since the effect would be 
small and  hard to separate from the enhanced emission from the magnetic 
elements.

Changes in surface heat flux by modified granulation, if they exist, will affect 
the convective  envelope in the same way as the excess emission from the small 
scale magnetic field. They can also be represented by an effective change of the 
emissivity of the solar surface. In particular, one does not expect these changes 
to be `compensated' by opposite changes elsewhere on the surface (except, as 
before, on the $10^5$yr thermal time scale of the envelope).

\section{Improving on the turbulent diffusion picture}
\label{new}
In the results quoted a diffusion model for convection was used. It assumes that 
convection can be modeled by a turbulent viscosity for momentum and a 
turbulent diffusivity for the transport of heat.  While this was a simple and 
somewhat justifiable model of convection in the absence of detailed knowledge 
of the convective flows in a stellar envelope, we now know that it does not 
represent stellar convective envelopes well. Numerical simulations
(Nordlund 1982, 1985ab, 1986, 1990, Nordlund and Dravins 1990, Nordlund and 
Stein 1990, 1991, 1996, Stein and Nordlund 1989, 1991, 1998, Steffen et al. 
1989, Steffen 1993, Ludwig et al. 1999) show an extremely {\it nonlocal} 
picture. The convective flows are driven almost exclusively by cooling at the 
surface, with narrow fast moving downdrafts between slow almost isentropic 
upflows (for a discussion see Spruit, 1997). In the present context, the most 
important property is that the convective flow at all depths is driven by cooling 
at the surface rather than by a local overturning process. The material in the 
cool downdrafts survives to large depths below the surface with little mixing 
into the upflows.

Given this extremely nonlocal picture, it is appropriate to ask how the thermal 
behavior of perturbations in the convection zone can be estimated without 
appealing to a turbulent diffusion model. I do this below for three of the effects 
discussed, the spot blocking, the facular emission, and the `thermal shadow' 
problems. In all three cases, the conclusion is that thermal perturbations of the 
normal solar surface are even smaller than in the diffusion picture. This has to 
with the fact that the flows in the new non-local picture are caused by cooling 
at the surface, and that most of the changes are one-way (downward, away from 
the visible surface). The diffusion picture on the other hand communicates 
perturbations equally well in all directions.

\subsection{spot blocking}
\label{spot:new}
Below the spot (modeled as a region of reduced heat transport efficiency 
extending to some depth below the surface) the upflows have exactly the 
same temperature as upflows in the unspotted surroundings at the same level, 
namely that given by the entropy at the base of the upflows. In this sense, there 
is no `pile up of heat below the spot'. Because of the reduced heat loss at the 
surface, however, the downflows below the spot will be less vigorous. The 
unspotted surface notices nothing of the spot's presence (except for an 
extremely narrow ring where lateral radiative exchange takes place, 
and except for the presence of a moat flow, see below). It continues to cool 
upwellings into downdrafts as before, since the entropy in the upflows has not 
changed. Thus, we expect again that bright rings will be absent around spots, 
but the reason is even simpler than in the diffusion model. The spot is a region 
at the surface where less heat is radiated away, and this is now independent of 
the depth of the spot below the surface (in contrast to the diffusion model, 
where the spot has to extend to a minimum depth of 1000 km for the 
explanation to work). 

Still, one may wonder what happens to the amount of heat generated in the solar 
interior that  now fails to be emitted at the surface. This part of the problem is 
the same as in the  diffusion model: the imbalance causes a secular increase of 
the entropy in the entire convection zone until a new thermal equilibrium is 
reached. Because of the very long thermal time scale of the convection zone 
($10^5$ yr), the effect is negligible on observable time scales. In a steady 
state, when the average number of spots does not change, the convection zone 
does not heat up, because its mean temperature is higher than it would be 
without spots (Spruit and Weiss, 1986). Episodes of larger than average spot 
coverage cause heating, those of less than  average spot coverage cause cooling 
on this time scale. 

\subsection{facular emission}
\label{new:fac}
The lower internal pressure in a small magnetic element (`tube') in the 
photosphere causes it to be more transparent: a surface of unit optical depth as 
seen from earth has a little dimple at the loaction of the small magnetic 
element. Its side walls radiate an additional heat flux (Spruit 1976, 1977). This 
heat 
flux is supplied to the element by convection, i.e. by the granulation surrounding 
it, and conversely the side wall, by the extra radiation it emits, cools the 
surroundings. In a simple minded diffusion model for convection, this causes a 
dark ring around the element, which compensates for a part of the extra 
emission from the element. The cooling, however, also makes the photosphere 
next to the tube heavier, so that a circulation is set up: down along the tube 
walls, and towards the element at the surface. This effect has been studied 
with numerical simulations by Kn\"olker et al. (1991, see also Steiner et al. 
1998), who find a very fast downflow along the tube wall (km/s) due to the 
excess cooling. In a diffusion model, the information of the low temperature in 
the downdrafts diffuses back to the surface. In reality, the cool downdraft 
drops below the visible surface because it is heavy, and only a negligible 
fraction of it finds its way back to the surface. The excess emission is 
therefore expected to be stronger than in the diffusion model, and not 
compensated by a dark ring.

\subsection{thermal shadows}
\label{shad:new}
The thermal shadow expected from a blocking object below the surface (Spruit, 
1977), which is a small effect already in the turbulent diffusion model, is 
likely to be even smaller in the new picture of stellar envelope convection. 
Since all upward flowing gas has nearly the same entropy, a blocking object 
below the surface has no effect on what happens above it, as long as there is 
enough room above it for upflows to comfortably reach the solar surface. It may 
interfere somewaht with downflowing gas, but this has no consequence for the 
surface temperature above the blocking object, since only a small fraction of 
the downward moving gas is carried back up to the surface. As before, this is a 
consequence of the non-local nature of the  convective flow.

\section{Helioseismological tests}
\label{tests}
In the above I have concentrated on surface effects, and argued that 
temperature anhancements with origins deep inside the convection zone are less 
likely to contribute. Such deep origins have been advocated by Kuhn et al. 
(1988), and Kuhn and Stein (1996) who propose that a temperature enhancement 
with origin at the base of the convection zone causes a temperature 
enhancement of 0.5K in the active latitudes. 

Potentially exhausing theoretical debates for or against these models can 
probably be circumvented since the models make different predictions for the 
associated perturbations below the surface that can be tested by observation. 
The excess radiation from small scale magnetic fields comes about through an 
increased surface emissivity, associated with the increased `surface 
roughness'. This increases the average cooling rate at the surface and creates 
somewhat stronger and/or cooler downdrafts. It therefore causes slightly {\em 
lower} horizontally averaged temperatures below the surface. The two models 
thus predict opposite thermal effects below the surface. 

The prospects for a test of these predictions are good, since helioseismological 
data are now quite sensitive, with the data obtained by SOHO/MDI and the GONG 
network (Kosovichev 1996, Kosovichev and Schou 1997, Duvall et al. 1998, 
Duvall and Kosovichev 1999). Changes in the p-mode frequencies related to the 
solar cycle were reported by e.g. by Libbrecht and Woodard (1990), Dziembowski 
et al. (1998), and Dziembowski (this volume). The dependence of the signal on 
wavenumbers $(l,m)$ shows that most of the effect is due to a change in wave 
propagation in a layer very close to the solar surface, at the latitudes of 
magnetic activity. The sense of the effect, an increase of the frequencies, is the 
{\em opposite} (Goldreich et al. 1991) of what would be expected if the 
temperature in these latitudes were higher. Though a higher temperature 
increases the propagation speed (increasing the frequencies), it also expands 
the solar envelope at these latitudes, by vertical hydrostatic balance. This 
lengthens the path traveled and decreases the mode frequencies. The net effect 
turns out to be dominated by the the second effect, the path length increase, 
because it is linear in $T$, while the propagation speed increases only as 
$T^{1/2}$. 

In addition to the thermal effect, the presence of the magnetic field 
also has an effect, since it increases the `stiffness' of the gas (in terms of 
propagation, the relevant speed is the fast mode speed, which is larger than the 
sound speed). Since the contribution of this effect scales with the ratio of 
magnetic to gas pressure, it is important only in a thin  layer near the surface. 
The observed changes in mode frequency can thus be due either to a somewhat 
lower temperature, or the effect of the magnetic field in the atmosphere. The 
observation that most of the effect is localized at the surface argues for a 
major contribution from the magnetic stiffness effect (Goldreich et al. 1991, 
Woodard et al. 1991). In any case, the observed sign of the effect is 
incompatible with enhanced temperatures being the main change in envelope 
structure in the active latitudes.

A second way of measuring variations in propagation conditions below the solar 
surface is through time-distance seismology (Duvall et al. 1996, Kosovichev 
1996, Braun et al. 1997), in which one does not measure the mode frequencies 
but the wave travel times between points on the surface (much like seismology 
is 
used to measure irregularities in the earth's interior). These measurements 
show shorter travel times in the active regions (Duvall et al. 1998, Duvall and 
Kosovichev 1998, Giles et al. 1997,  Bogdan et al. 1998, Goode and Dziembowski 
1999), and downflows below the surface. The shorter travel times are 
compatible with a contribution of the magnetic field to the propagation speed. 

The lower temperatures expected by enhanced surface cooling in the small scale 
magnetic field has further observational consequences.  In the absence of 
rotation, the increased downward buoyancy would just cause a downdraft below 
active regions. Associated with it a horizontal flow converging on the active 
region is expected. Effects like these are beginning to be explored by 
time-distance seismology.

\end{document}